\documentclass{JHEP3} 
\pdfoutput=1

\usepackage[utf8x]{inputenc}

\usepackage{graphicx, amsmath}

\newcommand{\be}{\begin{equation}}
\newcommand{\ee}{\end{equation}}
\newcommand{\bea}{\begin{eqnarray}}
\newcommand{\eea}{\end{eqnarray}}

\title{
Probing
Two Holographic Models \\
of Strongly Coupled Anisotropic Plasma}

\author{Anton Rebhan and Dominik Steineder\\
Institut f\"{u}r Theoretische Physik, 
Technische Universit\"{a}t Wien,\\
Wiedner Hauptstr. 8--10,
1040 Vienna, Austria\\ 

\\

 \email{rebhana@hep.itp.tuwien.ac.at}\\
\email{steineder@hep.itp.tuwien.ac.at}}

\abstract{Quark-gluon plasma
during its initial phase after its production in heavy-ion collisions
is expected to have substantial pressure anisotropies. In order
to model this situation by
a strongly coupled $\mathcal N=4$ super-Yang-Mills plasma with fixed anisotropy
by means of AdS/CFT duality, two models have been discussed in the
literature. Janik and Witaszczyk have considered a 
geometry involving a comparatively benign naked singularity,
while more recently Mateos and Trancanelli have used
a regular geometry involving a nontrivial axion field
dual to a parity-odd deformation of the gauge theory
by a spatially varying $\theta$ parameter.
We study the (rather different) implications of these two models on the
heavy-quark potential as well as jet quenching
and compare their respective predictions with
those of weakly coupled anisotropic plasmas.}


\begin{document}

\maketitle

 \section{Introduction}

In ultrarelativistic heavy-ion collisions, quark-gluon plasma is produced
far from equilibrium with strong anisotropy caused by the fact that initially
the system expands mainly along the collision axis. This complicates enormously
any theoretical analysis and makes it difficult to decide whether the strong
collectivity observed is indeed proving that the quark-gluon plasma behaves as a intrinsically strongly coupled, near-perfect fluid and that a description based on a perturbative plasma with collective effects from strong gluon fields can be ruled out \cite{Muller:2006ee}.

At weak coupling, an anisotropic quark-gluon plasma leads to (non-Abelian) plasma instabilities \cite{Mrowczynski:1988dz,Romatschke:2003ms,Arnold:2003rq,Arnold:2005vb,Rebhan:2005re,Rebhan:2008uj,Rebhan:2009ku,Attems:2012} leading to nonperturbatively large fields and turbulent behavior \cite{Arnold:2005ef,Berges:2008mr,Ipp:2010uy} which could be responsible for the strong collectivity. Using the framework of hard-loop effective theory \cite{Mrowczynski:2004kv}, experimental signatures such as anisotropic photon and dilepton emission \cite{Schenke:2006yp}, momentum broadening of jets \cite{Romatschke:2006bb,Baier:2008js} as well as heavy-quark potentials \cite{Dumitru:2007hy,Philipsen:2009wg} and quarkonium dissociation \cite{Burnier:2009yu} have been studied with fixed anisotropy as an approximation to the actual dynamical situation.

At strong coupling, for which conventional lattice gauge theory is of no help for analyzing strongly nonequilibrium dynamics, holographic gauge/gravity duality \cite{Aharony:1999ti,CasalderreySolana:2011us} offers the prospect of providing qualitative and semi-quantitative insights. In this framework, heavy-ion collisions and the subsequent anisotropic dynamics and eventual thermalization of a (super-)Yang-Mills plasma can be modeled e.g.\ by collisions of shock waves in anti-de Sitter space and horizon formation \cite{Chesler:2010bi}.

At least for certain observables, it may be meaningful to consider approximations with a temporarily fixed anisotropy in the gravity dual. One such attempt to model the effects of a system with anisotropic pressures was proposed by Janik and Witaszczyk in \cite{Janik:2008tc}, where the gravity dual involves a comparatively benign naked singularity. In \cite{Rebhan:2011ke} we have studied electromagnetic signatures of this model which are qualitatively similar to weak-coupling results at high frequencies.

More recently, Mateos and Trancanelli \cite{Mateos:2011ix,Mateos:2011tv} have proposed a regular gravity dual of an anisotropic but equilibrium $\mathcal N=4$ super-Yang-Mills plasma where a stationary anisotropy is introduced through a (parity-violating) deformation of the gauge theory with a $\theta$ parameter that depends linearly on one of the spatial coordinates. In contrast to the singular gravity dual of Ref.~\cite{Janik:2008tc}, this model has a hydrodynamical limit, and in \cite{Rebhan:2011vd} we have shown that it involves the remarkable feature that certain components of the viscosity tensor break the usual holographic bound of Einstein gravity duals. This model has been explored in \cite{Giataganas:2012zy,Chernicoff:2012iq,Chernicoff:2012gu,Fadafan:2012qu} with regard to the heavy quark potential, the drag force on quarks, and jet quenching. (Other holographic models of anisotropic fluids have been introduced and studied in Refs.~\cite{Erdmenger:2011tj,Gahramanov:2012wz,Ammon:2012qs}.)

The aim of the present paper is to compare the effect of anisotropy on the heavy quark potential and jet quenching in the two holographic models by Janik and Witaszczyk (JW) and Mateos and Trancanelli (MT) based on a singular gravity dual and on axion-dilaton gravity, respectively. These results are moreover compared with perturbative results for a weakly coupled anisotropic plasma for both prolate and oblate anisotropies.

The organization of this paper is as follows. 
After briefly reviewing in Sect.~\ref{sec:two} the two holographic models for an anisotropic super-Yang-Mills plasma of Refs.~\cite{Janik:2008tc,Mateos:2011ix,Mateos:2011tv}, in Sect.~\ref{sec:wc} we recall the effects of anisotropies on a weakly coupled plasma as described by hard anisotropic loop effective theory and we then consider a zero-coupling version of the MT model and its consequences. In Sect.~\ref{sec:hqp} we consider the heavy-quark potential in the two holographic models as well as in the two weak-coupling models of an anisotropic plasma. In the case of the hard anisotropic loops we reproduce results for the real part of the heavy quark potential obtained previously in Refs.~\cite{Dumitru:2007hy,Philipsen:2009wg}. In addition we note that in certain directions and at large distances the anisotropic heavy quark potential involves oscillatory behavior instead of an exponential tail, which is caused by electric plasma instabilities. Those are absent in the $\theta$-deformed weakly coupled gauge 
theory, which however also has a nonmonotonic behavior at large distances in the anisotropy direction. The holographic models first of all show a completely different large-distance behavior since there is a finite distance where the string connecting the heavy quark breaks. Below the string-breaking distance, the heavy quark potential in the JW model agrees with the hard anisotropic loop results in that for oblate anisotropies the binding is stronger along the anisotropy direction than transverse to it and also in that this feature is reversed for prolate anisotropy. However, in the MT model as well as in its zero-coupling version the potential is always deeper in transverse directions, for both prolate and oblate anisotropy. In Sect.~\ref{sec:jq} we compare the two holographic models with regard to jet quenching and momentum broadening parameters. Again we find that in the JW model the effects of anisotropy are opposite for prolate and oblate anisotropies, while the MT model is more uniform in this respect.
 However, in the case of anisotropic jet quenching parameters, the weak coupling results of Ref.~\cite{Romatschke:2006bb,Baier:2008js} show the opposite trend than those of the JW model while they happen to agree qualitatively for an oblate anisotropy in the case of the MT model. Sect.~\ref{sec:concl} contains our conclusions.

\section{Two holographic models of strongly coupled anisotropic plasma}\label{sec:two}

A primary measure of the anisotropy of the boundary field theory is the pressure anisotropy
\begin{align}
 \Delta=\frac{P_\perp}{P_z}-1,
\end{align}
where $\Delta>0$ ($\Delta<0$) corresponds to an oblate (prolate) plasma. In the following we shall recapitulate the main features of the two gravity duals we want to consider and also discuss the behavior of the stress-energy tensor in either case.

\subsection{JW model: singular gravity dual} 

The metric of the dual geometry given in Fefferman-Graham coordinates 
\begin{equation}
ds^2=\frac{\gamma_{\mu\nu}(x^\sigma,u)dx^\mu dx^\nu+du^2}{u^2},
\end{equation} 
can be related to the expectation value of the stress-energy tensor. Here $u$ is the holographic coordinate with $u=0$ defining the AdS boundary. The correspondence is such that near $u=0$ the metric should be given by  
\begin{equation}
\gamma_{\mu\nu}(x^\sigma,u)=\eta_{\mu\nu}+u^4 \gamma_{\mu\nu}^{(4)}(x^\sigma)+\mathcal{O}(u^6).
\end{equation}
with
\begin{equation}
\langle T_{\mu\nu}(x^\sigma)\rangle=\frac{N_c^2}{2\pi^2}\gamma_{\mu\nu}^{(4)}(x^\sigma).
\end{equation}
In \cite{Janik:2008tc} Janik and Witaszczyk fixed the (traceless) stress-energy tensor 
\be\langle T^{\mu}{}_{\nu}(x^\sigma)\rangle=\textnormal{diag}(\epsilon,P_\perp,P_\perp,P_z)
\ee 
and therefore the boundary conditions for the Einstein equations with a negative cosmological constant, without adding any further matter fields. The most general form of the metric respecting the symmetries of the stress-energy tensor is
\begin{equation}
ds^2=\frac{1}{u^2}\Bigl(-a(u)dt^2+c(u)\big(dx^2+dy^2\big)+b(u)dz^2+du^2\Bigr).
\label{eq:metricJW}
\end{equation}
Solving for the unknown functions $a(u)$, $b(u)$ and $c(u)$ one finds
\begin{eqnarray}
a(u)&=&(1+A^2 u^4)^{1/2-\sqrt{36-2B^2}/4}(1-A^2 u^4)^{1/2+\sqrt{36-2B^2}/4}\nonumber\\
b(u)&=&(1+A^2 u^4)^{1/2-B/3+\sqrt{36-2B^2}/12}(1-A^2 u^4)^{1/2+B/3-\sqrt{36-2B^2}/12}\\
c(u)&=&(1+A^2 u^4)^{1/2+B/6+\sqrt{36-2B^2}/12}(1-A^2 u^4)^{1/2-B/6-\sqrt{36-2B^2}/12},\nonumber
\end{eqnarray}
where $A$ and $B$ are related to the energy density and the pressures by
\begin{eqnarray}
\epsilon&=&\frac{N_c^2}{2\pi^2}\Big(\frac{A^2}{2}\sqrt{36-2B^2}\Big)\\
P_\perp&=&\frac{N_c^2}{2\pi^2}\Big(\frac{A^2}{6}\sqrt{36-2B^2}+\frac{A^2 B}{3}\Big)\\
P_z&=&\frac{N_c^2}{2\pi^2}\Big(\frac{A^2}{6}\sqrt{36-2B^2}-\frac{2A^2 B}{3}\Big)
\end{eqnarray}

$A$ is a dimensionful parameter which in the isotropic case ($B=0$) is related to temperature $T$ according to $A=\pi^2T^2/2$. Nonvanishing values of the dimensionless parameter $B$ characterize the anisotropy of the system, since $P_z$ and $P_\perp$ depend differently on $B$. For negative $B$ the plasma is prolate, while it is oblate for positive $B$.
In the following we shall consider in particular the values $B=\sqrt{2}$,
where $P_z=0$ ($\Delta=\infty$), and $B=-\sqrt{6}$ where $P_\perp=0$ ($\Delta=-1$).
In a plasma made of free particles, such values correspond to maximal anisotropies, but the above geometry permits also negative values of pressure components
for larger $B$ (limited only by $|B|<\sqrt{18}$).

The gravity dual we just described is pathological in the sense that a naked singularity appears whenever $B$ does not vanish. The singularity is however benign in the sense that one can still choose infalling boundary conditions at the singularity, such that the calculation of for example retarded current-current correlators is still possible \cite{Rebhan:2011ke}. It turns out that these current-current correlators show no hydrodynamical behavior and strongly deviate from the isotropic result even for arbitrarily small $B$'s for sufficiently small frequencies\footnote{A similar behavior has in fact been found recently for the out-of-equilibrium production rate of dileptons modeled by collapsing shells in AdS \cite{Baier:2012tc}.}. However, for larger frequencies the results are well behaved and smoothly approach the isotropic result for decreasing $|B|$. In \cite{Rebhan:2011ke} we therefore assumed that this background can approximately describe the non-equilibrium physics on short enough time scales. 
Eventually the plasma will evolve towards equilibrium and therefore stationarity is not a valid assumption anymore. Asking questions about zero frequency limits, which strictly speaking probe an infinite time span are not meaningful in this model. 
In principle, this is also a problem for the potential between two heavy quarks in the plasma which we shall compute below, but we expect that the stationary approximation can still provide some qualitative insight.

\subsection{MT model: axion-dilaton-gravity dual}

In \cite{Mateos:2011ix, Mateos:2011tv} Mateos and Trancanelli presented a completely regular and well behaved gravity dual to an anisotropic but static plasma. Their model is based on the spatially anisotropic duals of Lifshitz-like fixed point of \cite{Azeyanagi:2009pr}, but with AdS boundary conditions. This provides an anisotropic version of an $\mathcal{N}=4$ super-Yang-Mills plasma where the anisotropy is kept fixed by a parity violating deformation of the gauge theory
\begin{align}
 S_{gauge}=S_{\mathcal{N}=4}+\delta S, \qquad \delta S=\frac{1}{8\pi^2}\int\theta(z)\textnormal{Tr }F \wedge F,
\label{eq:deformation}
\end{align}
where $\theta(z)=2\pi n_{D7} z$ depends linearly on one spatial dimension. $n_{D7}$ is a constant with dimensions of energy and can be interpreted as a density of $D7$ branes. The complexified coupling constant of the SYM theory is related to the axion-dilaton of type \MakeUppercase{\romannumeral 2}B supergravity by
\begin{align}
 \tau=\frac{\theta}{2\pi}+\frac{4\pi i}{g_{YM}^2}=\chi+i e^{-\phi}.
\end{align}
For the deformation in (\ref{eq:deformation}) the axion field $\chi=a z$ is position dependent with $a=\lambda n_{D7}/4\pi N_c$ written in terms of the 't Hooft coupling $\lambda=g_{YM}^2 N_c$. Since the axion is magnetically sourced by $D7$ branes, the solution can be considered as a number of such branes dissolved in the geometry. It is important to note that since the $D7$ branes do not extend in the holographic direction, they do not reach the AdS boundary and therefore do not introduce new degrees of freedom in the $\mathcal{N}=4$ SYM theory. The brane setup is summarized in Table \ref{tab:brane}.

\begin{table}
\begin{center}
\begin{tabular}{cc||cccc|c|c}
 & & $t$ & $x$ & $y$ & $z$ & $u$ & $S^5$ \\
\hline
$N_c$ & $D3$ & x & x & x & x & & \\
$n_{D7}$ & $D7$ & x & x & x & & & x \\

\end{tabular}
 \end{center}
\caption{Brane set up.}
\label{tab:brane}
\end{table}

The solution for the ten-dimensional bulk geometry that we are eventually interested in is a direct product of a five-dimensional manifold $\mathcal{M}$ with a negative cosmological constant $\Lambda=-6/L^2$ and $S^5$ with radius $L$ given by $L^4=4\pi g_s N_c l_s^4$. Therefore it suffices to consider the five dimensional axion-dilaton-gravity action 
\begin{align}
 S_{bulk}=\frac{1}{2\kappa^2}\int_{\mathcal{M}}\sqrt{-g}\Big(R+12-\frac{(\partial \phi)^2}{2}-e^{2\phi}\frac{(\partial \chi)^2}{2}\Big)+\frac{1}{2\kappa^2}\int_{\partial \mathcal{M}} \sqrt{-\gamma}2K,
\label{eq:action bulk}
\end{align}
   where $\kappa^2=8\pi G=4\pi^2/N_c^2$ and $L=1$. The line element in the string frame is of the form\footnote{Here $u$ does not correspond to the holographic coordinate in Fefferman-Graham coordinates as before.}
\begin{align}
 ds^2=\frac{1}{u^2}\Big(-\mathcal{F}(u)\mathcal{B}(u)dt^2+dx^2+dy^2+\mathcal{H}(u)dz^2+\frac{du^2}{\mathcal{F}(u)}\Big).
 \label{eq:metricMT}
\end{align}
 In the following we will stop to write the dependence on the holographic coordinate $u$ explicitly. We note that reparametrization invariance is already used to fix the coefficient in front of $dx^2$ and $dy^2$ and that $\mathcal{B}$ cannot be set to unity in general. If $\mathcal{H}=1$ we would get an isotropic solution. $\mathcal{F}$ is the blackening factor that must vanish at the position of the horizon $u=u_h$. It turns out that all the functions $\mathcal{B},\mathcal{F}$ and $\mathcal{H}$ can be written in terms of the dilaton $\phi$, which itself has to satisfy a third order nonlinear differential equation in $u$.

We note that the temperature and the entropy density are well defined since the solution under consideration is static. The temperature can be found from the regularity condition on the metric after Euclidean continuation and is given by $T=|\mathcal{F}'(u_h)|\sqrt{B_h}/4\pi$. The entropy density is given by a quarter of the horizon area over spatial volume.

The thermodynamics of this setup is discussed in detail in \cite{Mateos:2011tv}. To summarize some of the most important points:
\begin{itemize}
 \item Holographic renormalization brings in a reference scale $\mu$ and therefore the stress-energy tensor of the boundary theory shows a conformal anomaly $\langle T^\mu_\mu\rangle \propto a^4$. The energy density and the pressures depend separately on $T/\mu$ and $a/\mu$.
 \item When we keep the temperature constant and increase the anisotropy parameter from the isotropic limit $a=0$ the pressure anisotropy first always becomes oblate. The maximal value of $\Delta$ depends crucially on the temperature. After this initial oblate phase there always exists a special value for $a$ where the pressures in transverse and longitudinal direction coincide (without the bulk geometry becoming isotropic) and if $a$ is increased further the plasma becomes increasingly prolate. However regardless of the pressure anisotropy, the bulk geometry of the MT model is uniformly prolate, $\mathcal H\ge 1$.
 \item For small values of $a$ the plasma is unstable against filamentation along the $z$-direction. It is thermodynamically favorable to have regions in $z$ (but infinitely extended in $x$- and $y$-direction) that are isotropic and regions with a larger value of $a$. However, the interval of $a$'s for which these filamentation instabilities are present is smaller than the interval for the oblate plasma. In other words, the prolate phase is always stable, but there also exist oblate and stable phases.
\end{itemize}

\begin{table}
\begin{center}
\begin{tabular}{c|c||c|c|c|c|c}
 $\tilde\phi_h$ & $u_h$ & $a/T$ & $T$ & $N_c^{-2}s$ & $\Delta$ & $\epsilon/\epsilon_{iso}$\\
\hline
$-\infty$ & 1 & 0 & 0.318 & 0.159 & 0 & 1 \\
$-21/40$ & 201/200 & 1.32 &0.318 & 0.163 & 0.08 & 1.01 \\
3/50 & 107/100 & 6.43 & 0.318 & 0.201 & $-1.00$ & 1.69 \\
48/625 & 7/5 & 50.51 &0.318 & 0.383 & $-1.29$ &$9.42\cdot10^3$ 
\end{tabular}
 \end{center}
\caption{Choice of parameters and corresponding thermodynamic quantities for approximately constant temperature in the anisotropic axion-dilaton gravity dual.}
\label{tab:temperature}
\end{table}

\begin{table}
\begin{center}
\begin{tabular}{c|c||c|c|c|c|c}
 $\tilde\phi_h$ & $u_h$ & $a N_c^{2/3}/s^{1/3}$ & $T$ & $N_c^{-2}s$ & $\Delta$ & $\epsilon/\epsilon_{iso}$\\
\hline
$-\infty$ & 1 & 0 & 0.318 & 0.159 & 0 & 1 \\
$-17/50$ & 41/40 & 1.13 & 0.314 & 0.159 & 0.18 & 0.97 \\
9/250 & 6/5 & 4.23 & 0.289 & 0.159 & $-1.13$ & 1.79 \\
$-619/5000$ & 2 & 27.37 &0.231 & 0.159 & $-1.29$ &$6.62\cdot10^3$ 
\end{tabular}
 \end{center}
\caption{Choice of parameters and corresponding thermodynamic quantities for approximately constant entropy density in the anisotropic axion-dilaton gravity dual.}
\label{tab:entropy}
\end{table}

In the following sections we will compute the heavy quark potential and and the jet quenching parameter at constant temperature and at constant entropy density. In Table \ref{tab:temperature} and Table \ref{tab:entropy} we present our choice of parameters for these two situations and some related thermodynamic quantities (in units with $\mu=1$)\footnote{The parameter $\tilde\phi_h$ is related to the dilaton and the anisotropy parameter by $\tilde\phi_h=\phi(u_h)+\frac{4}{7}\log a$.}. These parameters are chosen so as to be in the same ballpark as those considered in Ref.~\cite{Mateos:2011ix,Mateos:2011tv}. With different parameters, also larger positive values of $\Delta$ are possible, however the following results do not change qualitatively.

\section{Two weak coupling models of a stationary anisotropic plasma}\label{sec:wc}

\subsection{Hard anisotropic loop effective theory}

\begin{table}
\begin{center}
\begin{tabular}{c||c|c|c|c|c}
$\xi$ & $\Delta$ & $N^{(n)}$ & $N^{(\epsilon)}$ & $p_{\rm hard}^{(n)}/T$
& $p_{\rm hard}^{(\epsilon)}/T$ \\
\hline
$-0.9$ & $-0.8365$ & 0.3162 & 0.1678 & 0.681 & 0.640 \\
0 & 0 & 1 & 1 &1 &1\\
10 & 6.442 & 3.317 & 4.075 &1.491 & 1.421 \\
100 & 55.72 & 10.05 & 12.74 &2.158 & 1.889\\
\end{tabular}
\end{center}
\caption{Anisotropy parameters in the hard anisotropic loop effective theory}
\label{tab:xi}
\end{table}

In a weakly coupled (nearly collisionless) ultrarelativistic gauge theory
plasma there is a hierarchy of scales, with hard scales $p$ defined as
typical energies and momenta of plasma constituents, and soft scales $gp$,
with coupling constant $g\ll1$,
pertaining to leading-order collective phenomena
such as Debye screening and plasmon masses. 
In thermal equilibrium, the effective
theory of soft scales is provided by the ``hard thermal loop'' effective
action \cite{Braaten:1992gm}. 
With anisotropic distribution functions for hard particles,
the corresponding ``hard anisotropic loop'' effective theory 
\cite{Mrowczynski:2004kv} involves
a rich spectrum of stable and unstable modes at momentum scales $gp$, which
have been worked out completely for axisymmetric
deformations of distribution functions
of the form \cite{Romatschke:2003ms}
\be\label{faniso}
f(\vec p)=N f_{\rm iso}(\sqrt{\vec p^2+\xi p_z^2}/p_{\rm hard})
\ee
with anisotropy direction $z$ and some normalization factor $N(\xi)$ with
$N(0)=1$ for vanishing anisotropy parameter $\xi$. 
A prolate momentum distribution is obtained for $-1<\xi<0$, whereas $\xi>0$
parametrizes oblate momentum distributions.

While the pressure anisotropy $\Delta$ is directly determined by $\xi$ (see Table \ref{tab:xi}), a comparison of quantities at different anisotropy is rather ambiguous \cite{Margotta:2011ta}. This could be done, e.g., by keeping the number density or the energy density fixed, but in both cases it also depends on whether this is done
by adjusting the normalization $N$ or the parameter $p_{\rm hard}$.
Keeping
number densities of hard particles fixed by adjusting $N$, as done in
Ref.~\cite{Philipsen:2009wg}, leads to $N^{(n)}(\xi)=\sqrt{1+\xi}$,
whereas constant energy density in hard particles requires
$N^{(\epsilon)}(\xi)=\mathcal R^{-1}(\xi)$
with 
\be
\mathcal{R}(\xi)=
\begin{cases}
\frac12\bigl[(1+\xi)^{-1}+\xi^{-1/2}
\text{arctan}(\sqrt{\xi})\bigr] &\text{for }\xi>0 \\ 
\frac12\bigl[(1-\xi)^{-1}+(-\xi)^{-1/2}
\text{atanh}(\sqrt{-\xi})\bigr]  &\text{for }\xi<0
\end{cases}
\ee

Alternatively, one could compare isotropic and anisotropic
plasmas by fixing $N=1$ and
adjusting $p_{\rm hard}$. Keeping the number density constant
requires
  \begin{eqnarray}
   p_{\rm hard}^{(n)}=(1+\xi)^{1/6}T,
  \end{eqnarray}
whereas for constant energy density one has
  \begin{eqnarray}
   p_{\rm hard}^{(\epsilon)}=\mathcal{R}^{-1/4}(\xi)T, 
  \end{eqnarray}
with $T=p_{\rm hard}|_{\xi=0}$. Following Refs.~\cite{Burnier:2009yu,Margotta:2011ta}, we shall mainly consider the option of rescaling $p_{\rm hard}$ in Sect.~\ref{compweak}.

At leading order, 
an anisotropic distribution function of the form (\ref{faniso})
gives rise to a polarization tensor of the form $\Pi_{\mu\nu}(k)=m_D^2 \hat\Pi_{\mu\nu}(\omega/|\vec k|,|k_z|/|\vec k|,\xi)$ with four independent dimensionless structure functions and $m_D^2$ the isotropic Debye mass.

For any nonzero $\xi$, the (chromo-)magnetostatic propagator turns out to involve
spacelike poles and unstable modes corresponding to a filamentation (or Weibel) instability. Below we shall be interested also in the (chromo-)electrostatic potential given by the Fourier transform of the electrostatic propagator 
\begin{equation}\label{D00hal}
D_{00}(\omega=0,\vec k)=\frac{\vec k^2+m_\alpha^2+m_\gamma^2}{(\vec k^2+m_\alpha^2+m_\gamma^2)(\vec k^2+m_\beta^2)-m_\delta^4}
\end{equation}
where $m_{\alpha,\beta,\gamma,\delta}^2$ are elementary but rather unwieldy functions of $\xi$ and $k_z^2/\vec k^2$ (for explicit expressions see \cite{Dumitru:2007hy}).
This electrostatic propagator has poles at purely imaginary wave vectors, corresponding to Debye screening, but additionally poles at a range of real wave vectors, which correspond to electric plasma instabilities. The latter arise for wave vectors within (outside) a $45^o$ cone about the direction of anisotropy for the oblate (prolate) case.

\subsection{Anisotropic Chern-Simons deformation of weakly coupled gauge theories}\label{sec:thetaqcd}

As an alternative model of a plasma with fixed anisotropy we consider the
zero-coupling limit of a gauge theory with the deformation (\ref{eq:deformation}) present in the MT model
, i.e.\ a gauge theory with
Lagrangian
\be\label{Lanisoqcd}
\mathcal L=-\frac14 F_{\mu\nu}^a F^{a\mu\nu}-\frac{a}4  \epsilon^{\mu\nu\rho z}
A_\mu^a F^a_{\nu\rho}.
\ee

This is
similar to axion electrodynamics with constant spacelike axion gradient \cite{Ni:1977zz,Ralston:1995rt}\footnote{With timelike axion gradient, this model is known as the Carroll-Field-Jackiw model \cite{Carroll:1989vb} of a Lorentz and parity violating (but isotropic) modification of electrodynamics.} but in the spirit of the MT model we interpret $a$, which has dimension of inverse length, thermodynamically as a density of some conserved charge distributed along the spatial $z$ direction.

The Lagrangian (\ref{Lanisoqcd}) implies an anisotropic dispersion law for
gauge boson modes with two gauge-invariant
branches \cite{Ralston:1995rt,GRS}
\begin{equation}
\omega_\pm^2=\vec k^2+\frac{a^2}2\left(
1\pm \sqrt{1+\frac{4 k_3^2}{a^2}} \right).
\end{equation}
Since $\omega_\pm^2\ge0$, there are no tachyonic modes,
in contrast to the case of a timelike axion gradient \cite{Carroll:1989vb}
and also in contrast to the hard anisotropic loop effective theory.

Introducing a finite temperature, the
free energy per gauge boson turns out to be given by
\begin{equation}
\mathcal F^{(0)}=-\frac{\pi^2 T^4}{45}+\frac{a^2T^2}{48}-\frac{a^3T}{64}+O(a^4)
\end{equation}
in the limit $T\gg a$ \cite{GRS}.
In the high-temperature limit we can ignore renormalization ambiguities, which
appear in the $O(a^4)$ terms in the present case of zero coupling (a more complete discussion will be given in \cite{GRS}).
With the interpretation of $a$ as a density of a conserved charge (as in \cite{Mateos:2011tv}) which can be increased by compressing the volume such that $a$ scales inversely to its longitudinal extent, we
obtain the pressure anisotropy
\begin{equation}
P_z^{(0)}-P_\perp^{(0)}=a \frac{\partial \mathcal F}{\partial a}=
\frac{a^2 T^2}{24}-\frac{3a^3T}{64}+O(a^4)
\end{equation}
which indicates a prolate pressure anisotropy at sufficiently large $T/a$.
This is to be contrasted with the holographic infinite-coupling result 
of the MT model \cite{Mateos:2011tv}
\begin{equation}
P_z-P_\perp= 
-\frac{a^2 T^2}{16}+O(a^4)
\end{equation}
which corresponds to an {\em oblate} anisotropy when $a\ll T$.\footnote{Outside of the high-temperature limit, oblate as well as prolate phases appear, both in the strong-coupling MT model and in the zero-coupling case \cite{GRS}.}

In the following we shall compare the various models in particular
with regard to anisotropies in the heavy quark potential. In the
weak-coupling model given by (\ref{Lanisoqcd}), this is
obtained from the
anisotropic electrostatic propagator which has the simple form 
\begin{equation}\label{D00thetaqcd}
D_{00}(\omega=0,\vec k)=\frac1{\vec k^2+a^2(1-k_3^2/\vec k^2)}
\end{equation}
This corresponds to anisotropic Debye screening, without the complication
of electric instabilities which are present in the electrostatic propagator
of the hard anisotropic loop effective theory.
(In fact, there are also no magnetic instabilities in the $a$-deformed
magnetostatic propagator, although as will be discussed in \cite{GRS}
the phase diagram in the $a$-$T$-plane has regions of metastability
and absolute instability against filamentation towards 
inhomogeneous densities $a$ along the direction of anisotropy.)

 \section{Heavy Quark Potential}\label{sec:hqp}

\subsection{Holographic calculations}

We begin by discussing the heavy quark potential obtained from the Wilson-Polyakov loop which is dual to a fundamental string with spacelike separated endpoints at the AdS boundary \cite{Maldacena:1998im,Rey:1998bq,Brandhuber:1998bs} for the generic form of a metric describing stationary but spatially anisotropic geometries 
\begin{align}
 ds^2=g_{tt}(u)^2 dt^2+g_{xx}(u)\big(dx^2+dy^2\big)+g_{zz}(u)dz^2+g_{uu}(u)du^2.
\label{eq:metric}
\end{align}

The action for the hanging string is 
\begin{align}
 S=-\frac{1}{2\pi\alpha'}\int d^2\sigma \sqrt{-h},
\end{align}
with $h_{ab}=g_{AB}\partial_a X^M\partial_b X^N$ being the induced metric on the worldsheet. Here the indices $a, b$ are either $0$ or $1$ and $M, N=\{t,x,y,z,u\}$. Due to the symmetry in the transverse plane we can always choose a coordinate system such that the $y$-coordinate vanishes. Parametrizing the string worldsheet by $t$ and $u$ and making a stationary ansatz for $x=x(u)$ and $z=z(u)$, we obtain
\begin{align}
 S&=-\frac{1}{2\pi\alpha'}\int \ dt \ du \ \mathcal{L}\big(x'(u),z'(u),u\big)\\
  &=-\frac{\mathcal{T}}{2\pi\alpha'}\int du \sqrt{-g_{tt}(u)\big(g_{uu}(u)+g_{xx}(u)x'^2(u)+g_{zz}(u)z'^2(u)\big)}\nonumber,
\label{eq:action pot}
\end{align}
where primes denote derivatives with respect to the holographic coordinate $u$ and $\mathcal{T}$ is a constant coming from the time integration. We need to find the string profile and therefore evaluate the equations of motion for $x(u)$ and $z(u)$, which are of the form
\begin{align}
-g_{tt}(u)g_{xx}(u)x'(u)&=\Pi_x \mathcal{L}\big(x'(u),z'(u),u\big), \\
-g_{tt}(u)g_{zz}(u)z'(u)&=\Pi_z \mathcal{L}\big(x'(u),z'(u),u\big), 
\end{align}
$\Pi_x$ and $\Pi_z$ being constants of motion. Disentangling the above equations we end up with
\begin{align}
 x'^2(u)&=-\frac{\Pi_x^2 g_{uu}(u)g_{tt}(u)g_{zz}(u)}{g_{xx}(u)\Big[\big(g_{tt}(u)g_{xx}(u)+\Pi_x^2\big)\big(g_{tt}(u)g_{zz}(u)+\Pi_z^2\big)-\Pi_x^2\Pi_z^2\Big]}\\
 z'^2(u)&=-\frac{\Pi_z^2 g_{uu}(u)g_{tt}(u)g_{xx}(u)}{g_{zz}(u)\Big[\big(g_{tt}(u)g_{xx}(u)+\Pi_x^2\big)\big(g_{tt}(u)g_{zz}(u)+\Pi_z^2\big)-\Pi_x^2\Pi_z^2\Big]}.
\end{align}

For a hanging string that connects two spatially separated points at the boundary we expect $x'^2(u)$ and $z'^2(u)$ to become negative for $u>u_0$. Since the numerator is manifestly positive (note that in our conventions $g_{tt}(u)$ is negative for Lorentzian signature) the denominator has to vanish at some point $u_0$ and then becomes negative for increasing values of $u$. This is also in line with the requirement that $du/dx=du/dz=0$ at the turning point $u_0$. Evaluating the zero in the common factor of the denominators eventually leads to a equation that can be written as
\begin{align}
 \frac{\Pi_x^2}{g_{xx}(u_0)}+\frac{\Pi_z^2}{g_{zz}(u_0)}=-g_{tt}(u_0)>0.
\end{align}
This is the defining equation of an ellipse and therefore
\begin{align}
 \Pi_x^2&=-g_{tt}(u_0)g_{xx}(u_0)\sin^2\phi, \\
 \Pi_z^2&=-g_{tt}(u_0)g_{zz}(u_0)\cos^2\phi.
\end{align}

$x'(u)$ and $z'(u)$ are completely determined by $u_0$ and the angle $\phi$. To obtain the functions $x(u)$ and $z(u)$ we can make one further choice, namely that both $x$ and $z$ vanish at the turning point $u_0$. It is then easy to find the distance between the two string endpoints
\begin{align}
 L=2\sqrt{x^2(0)+z^2(0)}
\end{align}
and the energy of the configuration
\begin{align}
 E^{reg.}=-\frac{S}{\mathcal{T}}-\frac{1}{\pi\alpha'}\int_0^{u_h} du\sqrt{-g_{tt}(u)g_{uu}(u)}.
\end{align}

To calculate the action above we integrate from $u=0$ to the turning point $u=u_0$ in (\ref{eq:action pot}), which covers only half of the string and therefore we have to multiply by two in order to obtain the full result. The last term above is the energy of two straight strings hanging from the boundary to the horizon at $u_h$ and is necessary to regularize the amount of energy of the hanging string. This also means that the connected configuration is energetically favored as long as $E^{reg.} < 0$. It can be checked easily that for an isotropic geometry with $g_{zz}(u)=g_{xx}(u)$ we recover the already well known expression for the heavy quark static potential. If in the anisotropic case we restrict to the simpler cases where the string endpoints are either separated exactly along the $z$ or $x$ direction the above equations simplify and we reproduce the same solutions as given previously in \cite{Giataganas:2012zy}.
Our expressions above however are valid for any separation of the string endpoints in the $xz$-plane. The generic situation allows us to probe the geometry by letting the string hang down in the bulk and study how it deforms as a function of $u$. 

\begin{figure}
\begin{minipage}[t]{0.48\textwidth}
\begin{center}
 \includegraphics[scale=1]{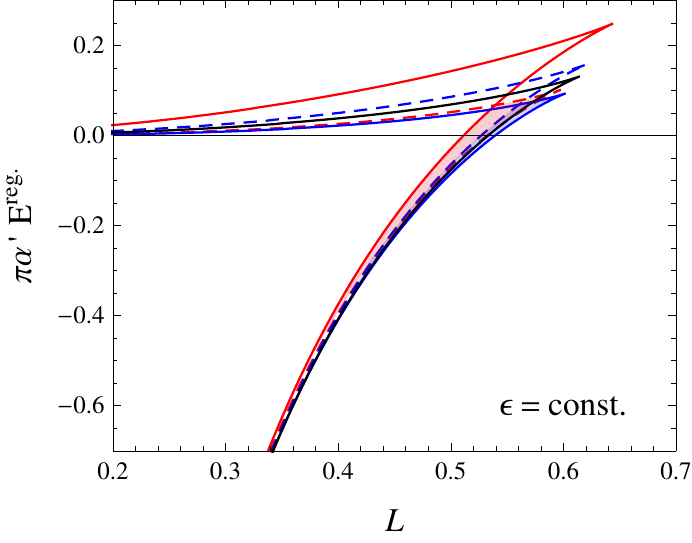}
\caption{Potential energy of heavy quarks in the JW model for plasmas with different anisotropies but constant energy density. The isotropic case corresponds to $B=0$ (black lines), oblate anisotropy with $P_z=0$ to $B=\sqrt{2}$ (blue) and prolate anisotropy with $P_\perp=0$ to $B=-\sqrt{6}$ (red). Full (dashed) lines refer to a separation of the quarks along (perpendicular to) the direction of the anisotropy.}
\label{fig:potEJW}
\end{center}
\end{minipage}
\quad
\begin{minipage}[t]{0.48\textwidth}
 \includegraphics[scale=1]{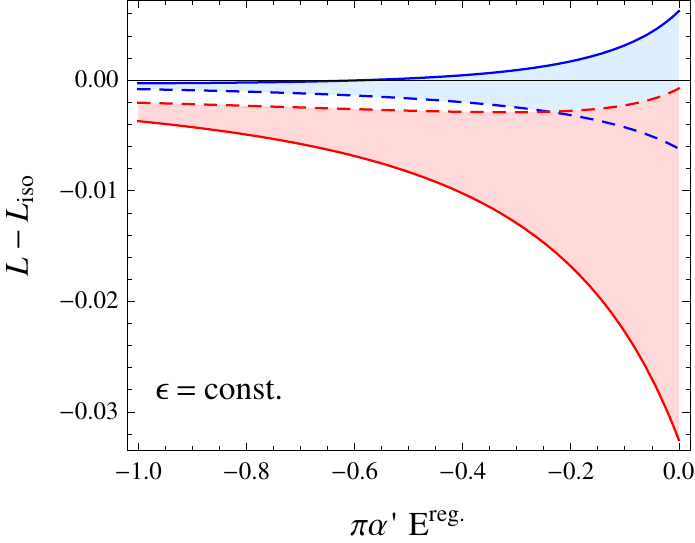}
\caption{Difference in the distance between two connected quarks at a given potential with its isotropic value for a plasma with oblate ($P_z=0$, $B=\sqrt{2}$, blue) and prolate ($P_\perp=0$, $B=-\sqrt{6}$, red) anisotropy at the same energy density. Full (dashed) lines  correspond to a separation of the quarks along (perpendicular to) the direction of the anisotropy.}
\label{fig:distEJW}
\end{minipage}
\end{figure}


We start by discussing the results for the JW model, the singular anisotropic gravity dual. In Fig.\ \ref{fig:potEJW} we plot the potential between the two heavy quarks, where we have adjusted the parameter $A$ of the model such that the energy density is kept constant for different anisotropies. 
Full (dashed) lines correspond to quarks separated along (transverse to) the direction of anisotropy. 

We note that in the oblate phase quarks separated along a transverse direction have a slightly shallower potential and consequently a smaller dissociation distance. (By dissociation distance we are referring to the maximal distance between two quarks, for which it is still energetically favorable to be connected by a hanging string in the bulk.\footnote{Strictly speaking, the dissociation of heavy quarkonia in a medium also depends on the imaginary part of the static potential in the real-time formalism which leads to a finite thermal decay width \cite{Laine:2006ns,Burnier:2009yu} and which we ignore by only studying Wilson loops in the Euclidean time direction.}) For prolate plasmas the heavy quark potential is instead shallower for longitudinally separated quarks than for transverse separations. 
Evidently the anisotropy only mildly influences the heavy quark potential even though we are considering extreme anisotropic plasmas with $P_z=0$ ($B=\sqrt{2}$) and $P_\perp=0$ ($B=-\sqrt{6}$). 
In Fig. \ref{fig:distEJW} we have made these small effects more conspicuous by plotting  the difference in the separation of two quarks at a given potential energy compared to the isotropic case. 



\begin{figure}
\begin{center}
 \includegraphics[scale=1]{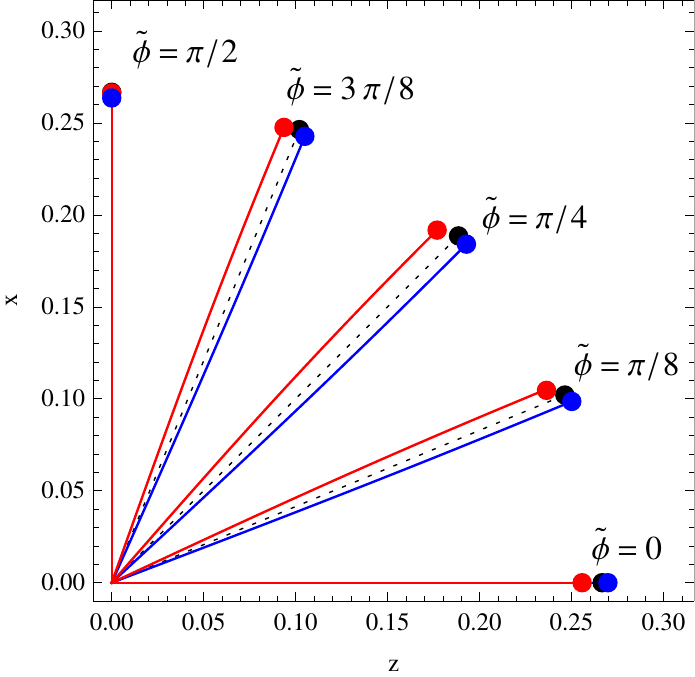}\qquad\includegraphics[scale=1]{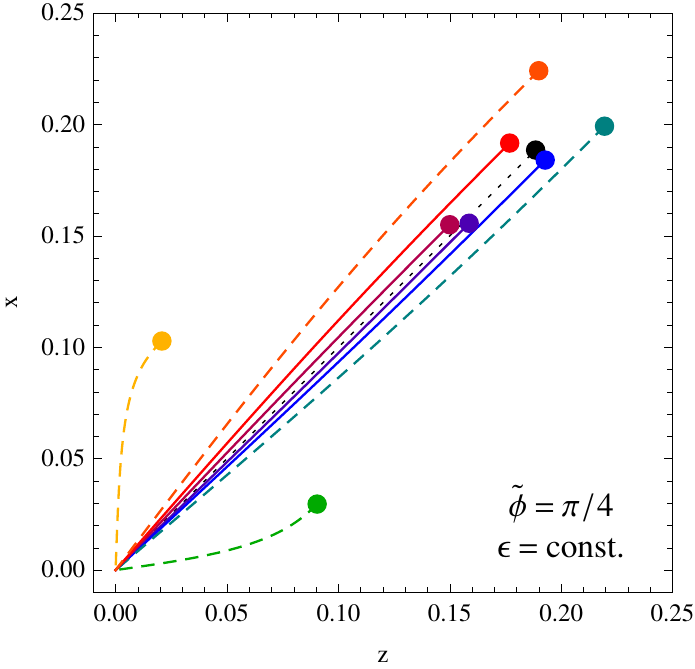}
\end{center}
\caption{String profiles in the JW model for boundary forces acting on the quarks pointing in a fixed direction (as indicated by $\tilde\phi$) and projected onto the boundary. The left panel shows profiles for strings at the point of string breaking ($E^{reg.} = 0$) for  $B=0$ (black, dotted), $B=\sqrt{2}$ (blue) and $B=-\sqrt{6}$ (red) with $\epsilon=const$. The right panel shows strings with $\tilde \phi=\pi/4$ but different turning points for the hanging string. For $B=\sqrt{2}$ ($B=-\sqrt{6}$) the colors going from red to yellow (blue to green) correspond to $u=\frac45 u_{sb}$, $u_{sb}$, $\frac65 u_{sb}$ and $2 u_{sb}$ with $u_{sb}$ the value at string breaking. Here dashed lines indicate unstable string configurations.}
\label{fig:stringsEJW}
\end{figure}

Let us finally
study the profile of the hanging string in the singular anisotropic geometry of the JW model in more detail. Due to the deformation of the spacetime as we go away from the boundary, the string projected onto the boundary will not be a straight line. The direction of the force acting on the string endpoint at the boundary can be defined by an angle
\begin{align}
 \tan\tilde{\phi}=\frac{\Pi_x}{\Pi_z}.
\end{align}
If $\tilde{\phi}=0$ then the force acts along the $z$-axis. One could now think of the following experiment. We act with forces pointing in a specified direction in the $xz$-plane on two heavy quarks that are initially close together. We choose the strength of the forces such that the heavy quarks slowly start to separate more until they dissociate. When we keep the direction of the forces fixed the whole time the quarks will however not follow a straight line along the force due to the deformation of the space in the holographic coordinate. Instead we observe the behavior shown in Fig.\ \ref{fig:stringsEJW}. We note that for the JW background 
the strings bend differently depending on the sign of the $B$ parameter. In the right panel of Fig.\ \ref{fig:stringsEJW} we consider a string endpoint with a force acting in $\tilde{\phi}=\pi/4$ direction and vary the depth of the turning point of the hanging string. Therefore we can probe the geometry up to a certain value of the holographic coordinate. As $u$ increases the deformation of the string gets stronger and stronger. For strings hanging almost down to the singularity we notice that the strings get deformed in such a way that they smoothly fit in the remaining space. When we take a look at the 
line element of the singular gravity dual we note that for $B>0$ the $z$-direction disappears while for $B<0$ the transverse directions vanish and the space degenerates into an infinite line as we go to the singularity. Therefore in the JW model the pressure anisotropy is encoded very directly in the geometry which is probed by the hanging string.

\begin{figure}
\begin{center}
 \includegraphics[scale=1]{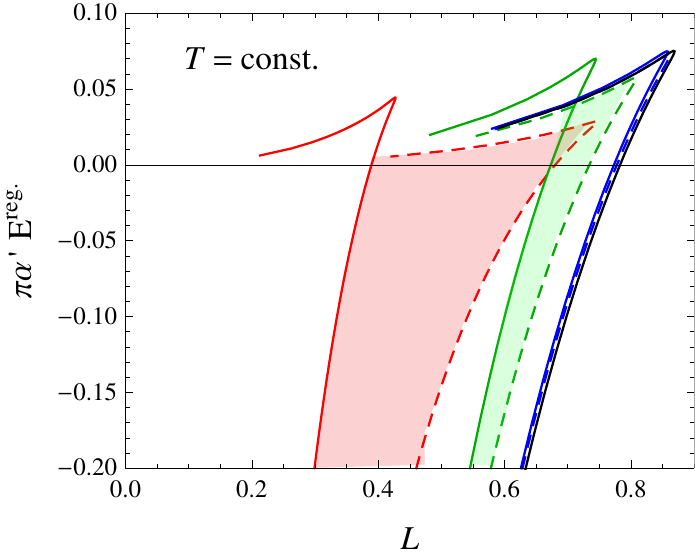}\qquad\includegraphics[scale=1]{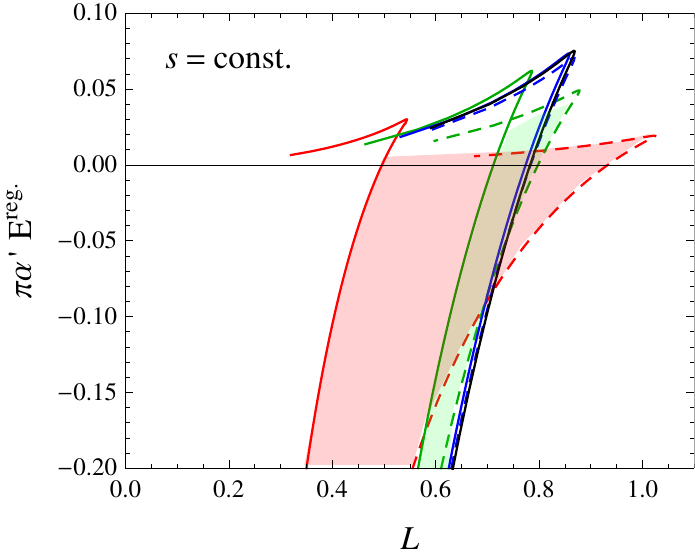}
\end{center}
\caption{Potential for heavy quarks in the MT model. The left panel compares varying anisotropies at constant temperature: $a/T=0$ (isotropic; black line), $a/T\approx1.32$ (oblate; blue lines), $a/T\approx6.43$ (prolate; green) and $a/T\approx50.51$ (prolate; red); the right panel at constant entropy density for different anisotropies $a N_c^{2/3}/s^{1/3}\approx1.13$ (oblate; blue), $a N_c^{2/3}/s^{1/3}\approx4.23$ (prolate; green) and $a N_c^{2/3}/s^{1/3}\approx27.37$ (prolate; red). Full (dashed) lines  correspond to a separation of the quarks along (perpendicular to) the direction of the anisotropy.}
\label{fig:potTMT}
\end{figure}


The anisotropic plasma of the MT model dual to axion-dilaton gravity is actually in thermal equilibrium and therefore we can compare the heavy quark static potential at constant temperature and at constant entropy density. The difference can be clearly seen in Fig.\ \ref{fig:potTMT}. 
At fixed temperature the dissociation length gets smaller for any separation in the $xz$-plane as we increase the anisotropy parameter $a$. At constant entropy density the difference of the dissociation length compared to an isotropic plasma depends on whether we separate the quarks along a transverse direction (string breaking occurs at a larger distance) or along the longitudinal direction (string breaking happens at a smaller distance compared to the isotropic result). 

However, regardless of the sign of the pressure anisotropy $\Delta$ 
we find that in the MT model the heavy quark potential is always deeper for transverse separation of the quarks.
(Note that the blue lines in the figures correspond to oblate configurations, while the green and red lines are for an increasingly prolate plasma, see Tables \ref{tab:temperature} and \ref{tab:entropy}). This is a striking difference to the situation in the JW model where oblate and prolate anisotropies lead to opposite deformations of the heavy quark potential. The situation in the MT model  is instead always similar to that in the JW model for prolate anisotropy. This appears to be a direct consequence of the fact that in the MT model $g_{zz}/g_{xx}=\mathcal H\ge 1$ for any $a$ whereas in the singular geometry of the JT model $g_{zz}/g_{xx}$ is larger (smaller) than unity for prolate (oblate) pressure anisotropy.

In the remaining plots we will only show the results for constant entropy density keeping in mind that for constant temperature the distance at which the string breaks becomes smaller and smaller as we increase $a$.

\begin{figure}
\begin{center}
 \includegraphics[scale=1]{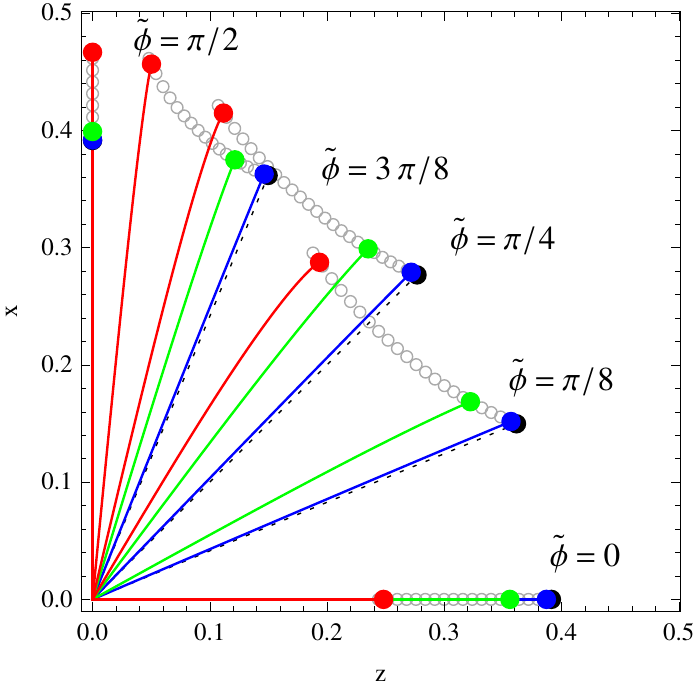}\qquad\includegraphics[scale=1]{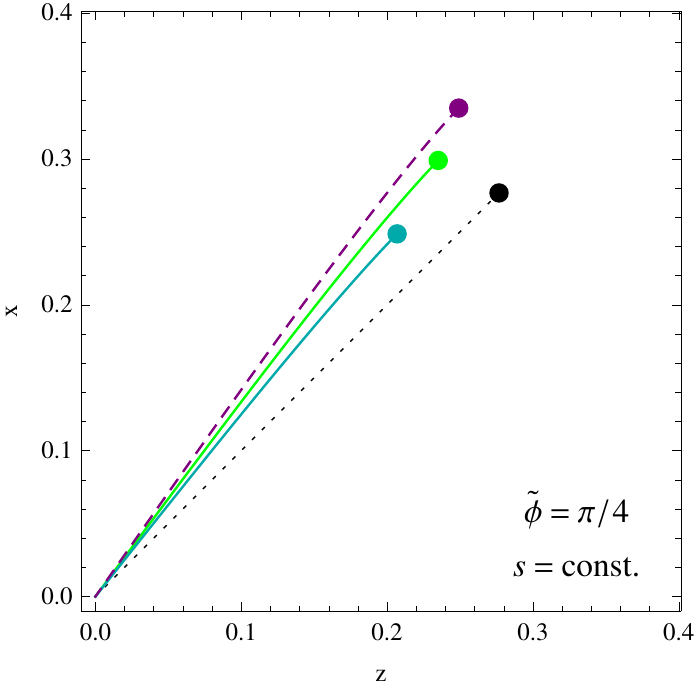}
\end{center}
\caption{String profiles in the MT model for boundary forces acting on the quarks pointing in the same direction (as indicated by $\tilde\phi$) and projected onto the boundary. The left panel shows the profiles for strings at the point of string breaking ($E^{reg.} = 0$) for $a N_c^{2/3}/s^{1/3}=0$ (black, dotted), $a N_c^{2/3}/s^{1/3}\approx1.13$ (blue), $a N_c^{2/3}/s^{1/3}\approx4.23$ (green) and $a N_c^{2/3}/s^{1/3}\approx27.37$ (red) with $N_c^{-2} s\approx0.159$. The right panel shows the profiles for $a N_c^{2/3}/s^{1/3}\approx4.23$ and $\tilde \phi=\pi/4$ but at different turning points for the hanging string: at string breaking $u_{sb}$ (green), $4 u_{sb}/5$ (cyan) and $6 u_{sb}/5$ (purple). Here dashed lines indicate unstable string configurations. }
\label{fig:stringssMT}
\end{figure}

Finally we also present the results for strings where the forces acting on the endpoints point in certain directions specified by the angle $\tilde{\phi}$. In Fig.\ \ref{fig:stringssMT} we also note that the situation is qualitatively the same as in the prolate case for the singular gravity dual. Also indicated in the plot are the trajectories the endpoint of the string follows as we increase the anisotropy and keep $\tilde{\phi}$ fixed. Here again we see once more that the geometry does change monotonically with increasing $a$ irrespectively of the behavior of the pressure anisotropy in the boundary theory. In the right panel of Fig.\ \ref{fig:stringssMT} we also probe the geometry by varying the location of the turning point of the hanging string. 

\begin{figure}
\begin{center}
 \includegraphics[scale=1]{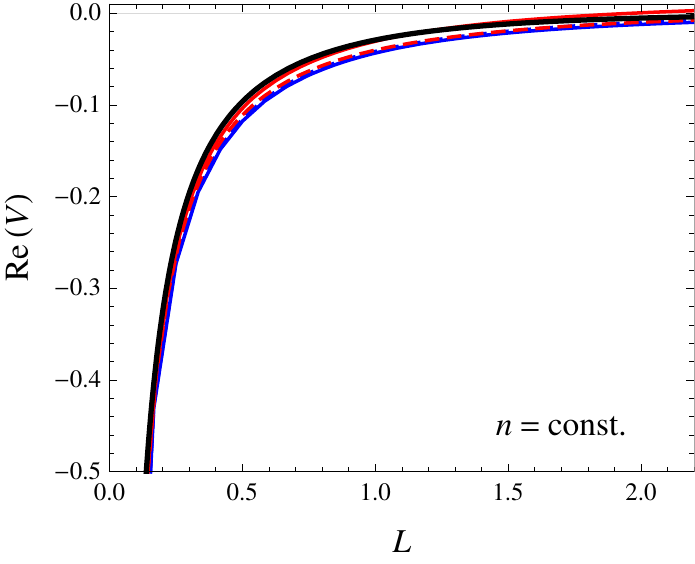}\qquad\includegraphics[scale=1]{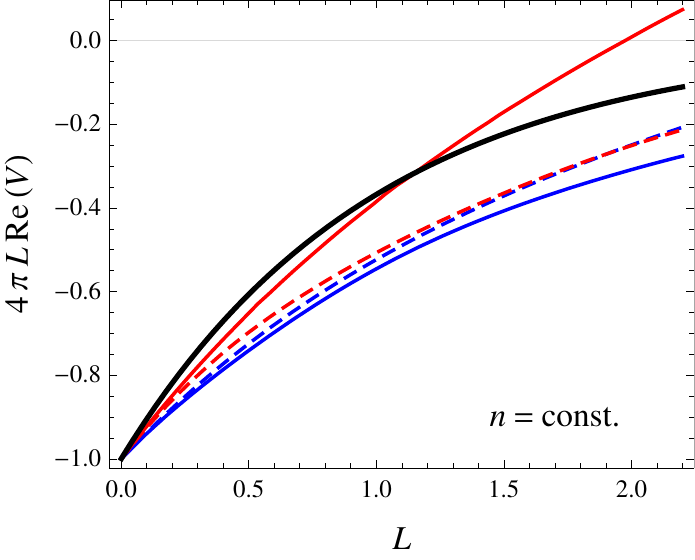}\\[10pt]
 \includegraphics[scale=1]{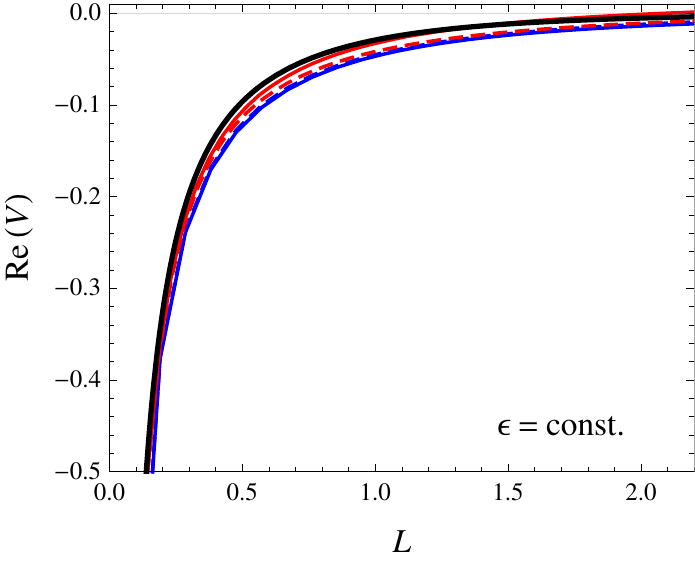}\qquad\includegraphics[scale=1]{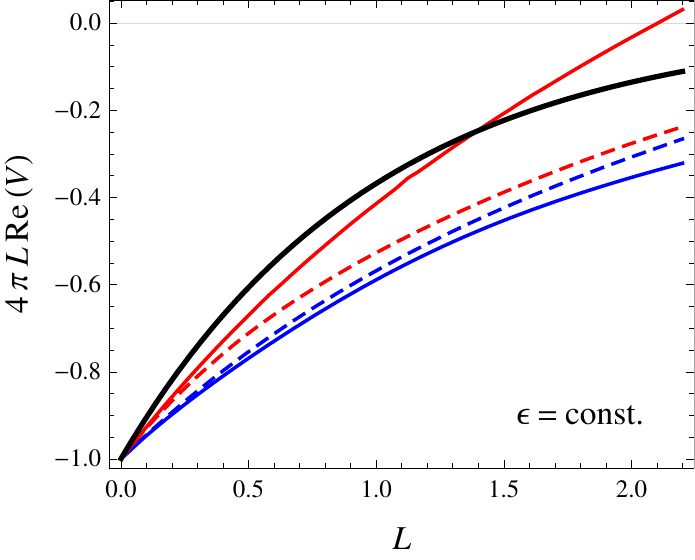}
\end{center}
\caption{Static potential for a weakly coupled anisotropic plasma in the hard anisotropic loop formalism as a function of the quark separation $L$ (units set by the isotropic Debye mass). The blue and red lines correspond to longitudinal (full) and transverse (dashed) orientation for $\xi=100$ and $\xi=-0.9$, respectively. The isotropic result is shown in black. In the upper two plots, the particle number density for the anisotropic and the isotropic plasma are the same, in the lower two plots the energy density is kept fixed.}
\label{fig:potweaks}
\end{figure}

\begin{figure}
\begin{center}
 \includegraphics[scale=0.9]{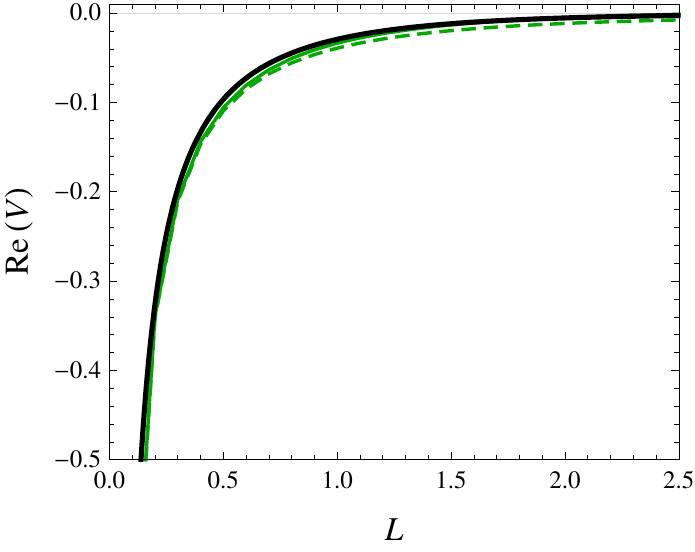}\qquad\includegraphics[scale=0.9]{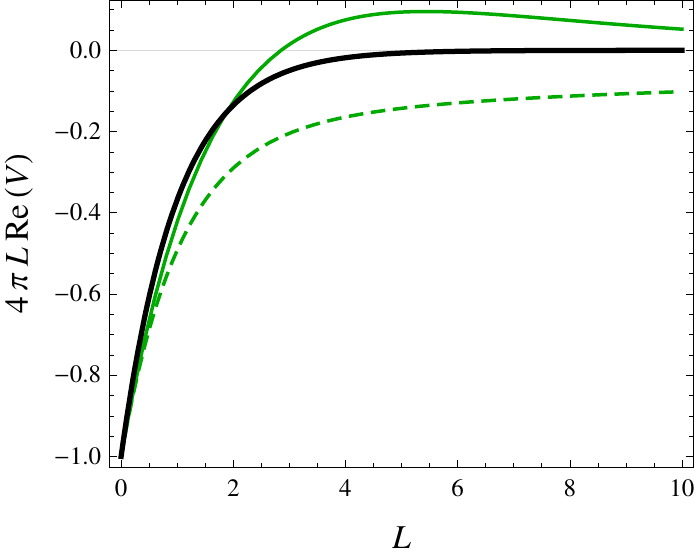}
\end{center}
\caption{Static potential in free anisotropic $\theta$-QCD for $a=1$. Full lines correspond to the $z$-direction and dashed lines to the $x$-direction.}
\label{fig:pottheta}
\end{figure}

\subsection{Comparison with weak-coupling calculations}\label{compweak}

At weak coupling, the real part of the heavy quark potential $V(\vec r)$
is given
by the Fourier transform of the electrostatic propagator. In
an axisymmetric situation 
integration over
the azimuth angle leads to
\be\label{Vweak}
V(\vec r)=-\frac1{4\pi^2}\int_0^\infty dk\int_{-1}^1 d\zeta\,
J_0(kr\sqrt{1-\zeta^2}\sin\theta_r)
\cos(kr\zeta\cos\theta_r)D_{00}(\omega=0,\vec k),
\ee
where $\cos\theta_r=z/r$ and $\zeta=k_z/k$ with our choice of the anisotropy direction
along $z$. In the case of the hard anisotropic loops, the propagator $D_{00}$
given by (\ref{D00hal}) involves poles at real $\vec k$ corresponding to electric plasma instabilities which are integrated over with a principal value prescription, while $D_{00}$ in the zero-coupling version of the MT model only
has poles at imaginary $\vec k$.

In Fig.\ \ref{fig:potweaks} we have evaluated (\ref{Vweak}) with the hard anisotropic loop propagator for strongly oblate ($\xi=100$) and prolate ($\xi=-0.9$) anisotropy (cf.\ Table \ref{tab:xi}), keeping alternatively the hard particle density $n$ and the energy density $\epsilon$ fixed for different anisotropies. The details of the deviation from the isotropic result slightly depend on whether $n$ or $\epsilon$ is kept constant, and in either case we find that for oblate anisotropies the heavy quark potential is slightly deeper along the anisotropy direction than transverse to it, while for prolate anisotropies this situation is reversed\footnote{This is also true when $N$ rather than $p_{\rm hard}$ is rescaled in (\ref{faniso}), but this method leads to somewhat stronger differences between fixed $n$ and fixed $\epsilon$.}. In order to make these effects more visible, we also plot $V(L)$ divided by the modulus of the vacuum (Coulomb) potential, $1/(4\pi L)$.

Comparing with the results of the JW model, we find a remarkable qualitative agreement in the dependence on the sign of the anisotropy and the direction of the quark separation. Moreover, the absolute deviation from the isotropic result is rather small both at weak coupling and in the JW model. 

On the other hand, as we have seen above, the MT model has a qualitatively different dependence on the direction of the quark separation in the case of  oblate anisotropies (which are usually considered in the context of heavy ion collisions). 

Turning to the zero-coupling version of the MT model introduced in Sect.\ \ref{sec:thetaqcd}, the heavy quark potential is given by the Fourier transform of 
(\ref{D00thetaqcd}) which is plotted in Fig.\ \ref{fig:pottheta}. As we have discussed in Sect.\ \ref{sec:thetaqcd}, the high-temperature limit of this weak coupling model corresponds to a prolate anisotropy (in contrast to the holographic MT model), whereas for general $a/T$ both prolate and oblate anisotropies are possible, depending on the renormalization scale. Curiously enough, the potential shown in Fig.\ \ref{fig:pottheta} (which does not depend on UV renormalization) has qualitatively similar dependence on the direction of quark separation as the holographic MT model (and the hard anisotropic loop potential in the prolate case).

We finally also consider the behavior of the quark potentials at large distances. In the two holographic models, there is a finite separation beyond which the string connecting the heavy quarks becomes unstable because strings entering the horizon or the naked singularity are energetically favored, and at a somewhat larger distance even no unstable connecting solution can be found.

To leading order at weak coupling, the isotropic quark potential is simply given by a Yukawa potential with exponential decay at large distance. The anisotropic weak coupling results show curious deviations. In the anisotropic $\theta$-deformed zero-coupling case Fig.\ \ref{fig:pottheta} shows a nonmonotonic behavior of the potential along the anisotropy direction such that beyond $L\sim 5$ (where the potential is actually already extremely small) there is even a repulsive behavior.

\begin{figure}
\begin{center}
 \includegraphics[scale=0.7]{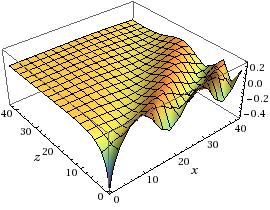}\qquad\includegraphics[scale=0.7]{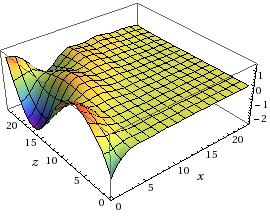}
\end{center}
\caption{Ratio of static potential to vacuum potential for $\xi=100$ and $\xi=-0.9$ at same energy density as in the isotropic plasma.}
\label{fig:oscillationsxz}
\end{figure}

Even more curious behavior can be found in the hard anisotropic loop potential at large distances. In this case there is nonmonotonic behavior along (transverse to) the anisotropy direction for prolate (oblate) anisotropy, and here the nonmonotonic behavior is moreover oscillatory. This behavior, which has not been noted in the previous studies of the hard anisotropic loop potential \cite{Dumitru:2007hy,Philipsen:2009wg}, is shown in Fig.~\ref{fig:oscillationsxz}, where the potential is plotted at large distances in the $xz$ plane (enhanced by dividing by the modulus of the vacuum (Coulomb) potential). This oscillatory behavior, which is reminiscent of Friedel oscillations at finite chemical potential (for a recent discussion see \cite{Blaizot:2005fd}), has its origin in the presence of poles in the electrostatic propagator at real wave vector corresponding to electric plasma instabilities. It is however rather clear that this curious behavior is devoid of physical implications even at weak coupling, because 
the plasma instabilities imply that a stationary anisotropy is only a justifiable approximation at sufficiently small time scales and correspondingly small length scales.

\section{Jet Quenching}\label{sec:jq}

\subsection{Holographic calculations}

The computation of the jet quenching parameter $\hat{q}$ for an anisotropic plasma with an axion-dilaton-gravity dual, the MT model, has been presented in \cite{Giataganas:2012zy, Chernicoff:2012gu}. Here we will reproduce the result for the most general case with an ultrarelativistic quark moving in an arbitrary direction \cite{Chernicoff:2012gu} and compare the results 
with those of the singular geometry of the JW model.

According to the prescription of \cite{Liu:2006ug, Liu:2006he} we calculate the string worldsheet with endpoints moving in the same direction at the speed of light and separated a small distance $l$ along a direction perpendicular to their motion. The jet quenching parameter $\hat{q}$ can then be obtained from
\begin{align}
 e^{2iS}=\langle W^A(\mathcal{C}_{lightlike})\rangle=\exp\Big(-\frac{L^- l^2}{4\sqrt{2}}\hat{q}\Big)+\mathcal{O}\Big(\frac{1}{N^2}\Big).\label{eq:qhat}
\end{align}

In the following we consider a quark endpoint moving in the $xz$-plane. The direction is parametrized by an angle $\theta$ such that for $\theta=0$ the quark moves along the $z$-axis. We therefore start by two subsequent coordinate transformations. First we define
\begin{align}
 Z&=z\cos\theta+x\sin\theta, \\
 X&=x\cos\theta-z\sin\theta, \\
 Y&=y
\end{align}
and then we introduce light-cone coordinates
\begin{align}
 Z^\pm=\frac{1}{\sqrt{2}}(t\pm Z).
\end{align}
The metric then takes the form
\begin{align}
 ds^2=& G_{++}(dZ^+)^2+G_{--}(dZ^-)^2+2G_{+-}dZ^+dZ^-\\
&+G_{XX}dX^2+2G_{X+}dXdZ^++2G_{X-}dXdZ^-+G_{YY}dY^2+G_{UU}dU^2\nonumber.
\end{align}
Writing the new metric coefficients in terms of our original ones we find\footnote{Here and in the following we will not write the dependence of the metric coefficients on the holographic coordinate $u$ in order to keep the expressions shorter. There is no danger of confusing at which value of $u$ we should evaluate the metric coefficient, because the string worldsheet in the calculation of the jet quenching parameter will always have its turning point at the horizon \cite{Chernicoff:2012gu}.}
\begin{align}
 G_{++}&=G_{--}=\frac{1}{2}\big(g_{tt}+g_{xx}\sin^2\theta+g_{zz}\cos^2\theta\big),\\
 G_{XX}&=g_{xx}\cos^2\theta+g_{zz}\sin^2\theta,\\
 G_{YY}&=g_{xx},\\
 G_{UU}&=g_{uu},\\
 G_{X+}&=-G_{X-}=\frac{1}{\sqrt{2}}\cos\theta\sin\theta(g_{xx}-g_{zz}).
\end{align}

We choose the worldsheet coordinates $(\tau,\sigma)=(Z^-,U)$ and let $Z^+$, $X$ and $Y$ depend on the holographic coordinate $U$ in the following. It is interesting that we must allow for a non-constant embedding of the string in $Z^+$ to find a solution in most general case. The Nambu-Goto action of the string is then given by
\begin{align}
 S=&-\frac{1}{2\pi\alpha'}\int d Z^- \int du \Big[G_{+-}^2 (Z^+)'^2+G_{X-}^2X'^2+2G_{+-}G_{X-}(Z^+)'X'\label{eq:action mb}\\
 &-G_{--}\big(G_{UU}+G_{++}(Z^+)'^2+G_{XX}X'^2+G_{YY}Y'^2+2G_{+X}(Z^+)'X'\big)\Big]^{\frac{1}{2}}\nonumber
\end{align}
The expression under the square root is actually negative which leads to an imaginary action. The reason is that we consider a spacelike string worldsheet. However this is expected because it is exactly what we need to obtain a jet quenching parameter that is real.

Since the Lagrangian does not depend on $Z^+$, $X$ or $Y$ explicitly we can find three constants of motion $\Pi_+$, $\Pi_x$ and $\Pi_y$. In the limit where these constants are small\footnote{We want to consider small separation lengths $l$ between the two string endpoints. In \cite{Chernicoff:2012gu} it is shown that this corresponds to the limit of small $\Pi$'s. As a further remark we note that the worldsheet turning point characterized by $dU/dX=0$ (and similarly for $Z^+$ and $y$) is located at the horizon.} we obtain
\begin{align}
 (Z^+)'&=c_{++}\Pi_+ +c_{+X}\Pi_X+\mathcal{O}(\Pi^2),\label{eq:Z+}\\
 X'&=c_{X+}\Pi_+ +c_{XX}\Pi_X+\mathcal{O}(\Pi^2),\label{eq:X}\\
 Y'&=c_{YY}\Pi_Y+\mathcal{O}(\Pi^2).\label{eq:Y}
\end{align}
In \cite{Chernicoff:2012gu} the coefficients $c$ are given explicitly for the metric of the axion-dilaton-gravity dual. Since we are interested in comparing the results of two different gravity duals we express these coefficients in terms of the general form of the metric given in (\ref{eq:metric}).
\begin{align}
 c_{++}=&\sqrt{\frac{g_{uu}}{2(g_{tt}+g_{zz}\cos^2\theta+g_{xx}\sin^2\theta)}}\ \frac{g_{tt}(g_{xx}\cos^2\theta+g_{zz}\sin^2\theta)+g_{xx}g_{zz}}{g_{tt}g_{xx}g_{zz}}\\
 c_{XX}=&\sqrt{\frac{2 g_{uu}}{g_{tt}+g_{zz}\cos^2\theta+g_{xx}\sin^2\theta}}\ \frac{g_{zz}\cos^2\theta+g_{xx}\sin^2\theta}{g_{xx}g_{zz}}\\
 c_{YY}=&\sqrt{\frac{2 g_{uu}}{g_{tt}+g_{zz}\cos^2\theta+g_{xx}\sin^2\theta}}\ \frac{1}{g_{xx}}\\
 c_{+X}=&c_{X+}=\sqrt{\frac{g_{uu}}{g_{tt}+g_{zz}\cos^2\theta+g_{xx}\sin^2\theta}}\ \frac{(g_{zz}-g_{xx})\sin\theta\cos\theta}{g_{xx}g_{zz}}
\end{align}
This agrees with \cite{Chernicoff:2012gu} if we insert the precise form of the metric in the axion-dilaton-gravity case. However it is now also straightforward to consider any background whose metric is of the form (\ref{eq:metric}).

The string endpoints at the boundary are not separated along the $Z^+$ direction and integrating (\ref{eq:Z+}) gives
\begin{align}
 \Pi_+=-\frac{\int_0^{u_h} du\ c_{+X}}{\int_0^{u_h}du\ c_{++}} \Pi_X.\label{eq:P+}
\end{align}
Along the $X$-axis the separation of the endpoints is $l\sin\phi$ while in $Y$-direction it is $l\cos\phi$. The constants of motion are then
\begin{align}
 \Pi_X&=\frac{l\sin\phi}{2}\frac{\int_0^{u_h} du\ c_{++}}{\int_0^{u_h} du\ c_{XX}\int_0^{u_h} du\ c_{++}-\big(\int_0^{u_h} du\ c_{+X}\big)^2},\label{eq:Px}\\
 \Pi_Y&=\frac{l\cos\phi}{2}\frac{1}{\int_0^{u_h} du\ c_{YY}}.\label{eq:Py}
\end{align}
If we insert the expressions (\ref{eq:Z+})-(\ref{eq:Y}) into the action (\ref{eq:action mb}) and expand to second order in $\Pi$'s we obtain
\begin{align}
 S=\frac{i L^-}{\pi\alpha'}\int_0^{u_h} du \sqrt{G_{--}G_{UU}}+\frac{i L^-}{2\pi\alpha'}\int_0^{u_h} du \big[c_{++}\Pi_+^2+c_{XX}\Pi_X^2+2c_{+X}\Pi_+\Pi_X+c_{YY}\Pi_Y^2\big].
\end{align}

The action is imaginary because we considered a spacelike worldsheet and $L^-$ is the length of the Wilson line in $Z_-$-direction. The first, $\Pi$ independent term is divergent, however, the jet quenching parameter is proportional to $l^2$ and therefore just contained in the second finite term. Upon inserting (\ref{eq:P+})-(\ref{eq:Py}) into the action and considering the defining relation for the jet quenching parameter (\ref{eq:qhat}) we eventually obtain
\begin{align}
 \hat{q}_{\theta,\phi}=\frac{\sqrt{2}}{\pi\alpha'}\big(P(\theta)\sin^2\phi+Q(\theta)\cos^2\phi\big)
\end{align}
 with
\begin{align}
 P(\theta)=&\frac{\int_0^{u_h} du\ c_{++}}{\int_0^{u_h} du\ c_{XX}\int_0^{u_h} du\ c_{++}-\big(\int_0^{u_h} du\ c_{+X}\big)^2},\\
 Q(\theta)=&\frac{1}{\int_0^{u_h} du\ c_{YY}}.
\end{align}

\begin{figure}
\begin{center}
 \includegraphics[scale=0.70]{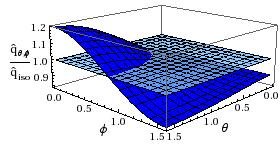}\qquad\includegraphics[scale=0.70]{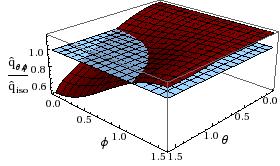}
\end{center}
\caption{Jet quenching parameter normalized to the isotropic result in the JW model. Left panel for an oblate plasma with $B=\sqrt{2}$, right panel for a prolate plasma with $B=-\sqrt{6}$, with $\epsilon=const$.}
\label{fig:MBJW}
\end{figure}

\begin{figure}
\begin{center}
 \includegraphics[scale=0.70]{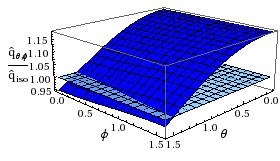}\qquad\includegraphics[scale=0.70]{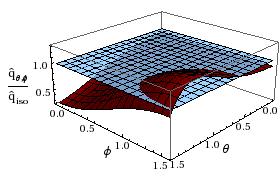}
\end{center}
\caption{Jet quenching parameter normalized to the isotropic result in the MT model. Left panel for a plasma at the same entropy density for $a N_c^{2/3}/s^{1/3}\approx1.13$ and $N_c^{-2}s\approx0.159$ which corresponds to oblate pressure anisotropy; right panel for prolate pressure anisotropy with $a N_c^{2/3}/s^{1/3}\approx27.37$.}
\label{fig:MBMT}
\end{figure}

The average 
\be
\hat{q}_\theta\equiv \frac1{2\pi}\int_0^{2\pi}\! d\phi\,\hat{q}_{\theta,\phi}
\equiv \frac12(\hat{q}_{\theta,0}+\hat{q}_{\theta,\pi/2})\equiv
\hat{q}_{\theta,\pi/4}
\ee
is the total jet quenching parameter for a quark moving with angle $\theta$ with respect to the anisotropy direction, while $\hat{q}_{\theta,\phi}$ contains the information about momentum broadening in directions transverse to the motion of the quark, with $\phi=0$ being perpendicular to both the direction of the jet and the anisotropy direction. When $\theta=0$, i.e.\ the quark moving along the direction of anisotropy, $\hat{q}_{0,\phi}\equiv \hat{q}_{0}$ is independent of $\phi$. In the context of heavy-ion collisions, one is of course mostly interested in jets at larger $\theta$. For $\theta=\pi/2$ one can define transverse and longitudinal jet quenching parameters
\be
\hat{q}_\perp = \hat{q}_{\pi/2,0}\,,\qquad
\hat{q}_L = \hat{q}_{\pi/2,\pi/2}\,.
\ee

In Figures \ref{fig:MBJW} and \ref{fig:MBMT} $\hat{q}_{\theta,\phi}$ is plotted for oblate and prolate pressure anisotropies in
the MT model and in the JW model.
Once more we see that in the JW model a different sign of the anisotropy parameter leads to a qualitative change of the result, in particular whether $\hat q_L$ is larger or smaller than $\hat q_\perp$, while the results for the MT model are more similar to the prolate phase of the JW model, for any parameter $a$ and thus regardless of the sign of the pressure anisotropy $\Delta$.

\subsection{Comparison with weak-coupling calculations}

Attempts to calculate the anisotropic jet quenching parameters for a quark moving transverse to the anisotropy direction ($\theta=\pi/2$) have been presented in Refs.~\cite{Romatschke:2006bb,Baier:2008js} based on one-loop calculations using hard anisotropic loops. In this situation the presence of spacelike poles in the static gluon propagator leads to nonintegrable singularities. In Ref.~\cite{Romatschke:2006bb} it was conjectured that these singularities might get be cured by the generation of an imaginary part in the static gluon self energy at higher loop order and this conjecture was used for an estimate of this ``anomalous'' contribution. While in Ref.~\cite{Carrington:2009vm} this conjecture was refuted, Ref.~\cite{Baier:2008js} proposed alternative resolutions, which all point to anomalous contributions of the same sign and angular dependence as the infrared-safe regular contributions computed previously 
in Ref.~\cite{Romatschke:2006bb} to leading logarithmic order. For a quark moving with the speed of light and in the limit of small anisotropy parameter $\xi$, the regular contribution to $\hat q_\perp$ and $\hat q_L$ was obtained as
\be
\hat{q}_{L,\perp}^{\rm reg.}=\hat{q}_{\rm iso}\left(1\pm \frac{\xi}{3}+O(\xi^2)\right)
\ee 
To linear order in $\xi$,
the anomalous contribution is in fact only present for oblate anisotropy\footnote{Refs.~\cite{Romatschke:2006bb,Baier:2008js} only discussed oblate anisotropies, which have $m_\alpha^2 \le 0$ and $m_\alpha^2\propto \xi$. For prolate anisotropy only the last term of Eq.~(32) in \cite{Baier:2008js} contributes, which is of order $\xi^3$.} ($\xi>0$) with
\be
\hat{q}_{L,\perp}^{\rm anom.}=C \hat{q}_{\rm iso}\left(\frac{\xi}{6}(1 \pm \frac12)\Theta(\xi)
+O(\xi^2)\right),
\ee 
where $C$ is a positive constant 
which depending on the physical cutoff for the singularities arising from plasma instabilities may differ from unity and also involve $\ln(1/\xi)$ \cite{Baier:2008js}. At any rate, to linear order in $\xi$ the hard anisotropic loop calculations of Refs.~\cite{Romatschke:2006bb,Baier:2008js} imply $\hat q_L>\hat q_\perp$
for oblate pressure anisotropy, and $\hat q_L<\hat q_\perp$ for the prolate case. This result neither agrees with the results of the JW model nor with those of the MT model: in the JW model the ordering of the two jet quenching parameter changes with the sign of the anisotropy, but the ordering is just the opposite. The MT model on the other hand always has $\hat q_L>\hat q_\perp$ which agrees with the hard anisotropic loop result in the oblate case, but differs in the prolate case.

It is actually questionable whether the one-loop calculation using hard anisotropic loops is relevant for the physics of a weakly coupled anisotropic quark-gluon plasma. Anisotropic jet quenching could be instead dominated by large chromomagnetic fields generated by plasma instabilities
\cite{Dumitru:2007rp,Majumder:2009cf}. Because plasma instabilities give rise to $|B_\perp|>|E_\perp|$ and $|E_L|>|B_L|$, it has been argued in Ref.~\cite{Dumitru:2007rp} that this would also give $\hat q_L>\hat q_\perp$ for a plasma with oblate anisotropy. In fact, the (different) plasma instabilities for prolate anisotropies equally lead to\footnote{This holds true for both weak fields \cite{Rebhan:2009ku} and nonperturbatively large fields \cite{Ipppriv}.} $|B_\perp|>|E_\perp|$ and $|E_L|>|B_L|$, thus also favoring $\hat q_L>\hat q_\perp$. Perhaps fortuitously, this is in line with the results of the MT model, although the latter of course neglects any dynamics from instabilities and the formation of inhomogeneous configurations\footnote{Note that the fields associated with plasma instabilities at weak coupling have nonvanishing wave number.}.

\section{Conclusion}\label{sec:concl}

In this paper we have studied two different holographic models of a strongly coupled super-Yang-Mills plasma with (temporarily) fixed pressure anisotropy. To this end, the JW model uses a singular anisotropic geometry with a naked singularity, whereas the MT model achieves a completely regular geometry and thus thermal equilibrium through an axion field with constant spacelike gradient which is dual to a parity-violating deformation of the gauge theory by a spatially varying $\theta$ angle. We have probed the different geometries by Wilson loops which in their respective forms are dual to the static potential of heavy quarks and jet quenching parameters.

In the case of the heavy quark potential, we have found that the results of the JW model agree qualitatively with a weak coupling description through anisotropic hard loops in that quarks separated moderately along the anisotropy direction have a slightly deeper (shallower) potential for oblate (prolate) pressure anisotropy. The MT model differs from the JW model in that the anisotropic deformation of the heavy quark potential is always similar to the prolate case of the JW model or the hard anisotropic loops, regardless of the sign of the pressure anisotropy. A similar result was found for a $\theta$-deformed weakly coupled gauge theory which we have introduced as an alternative model for an anisotropic (and parity-violating) plasma. 

At larger distances in the two holographic models the quarks break up, but the dissociation lengths are ordered differently with respect to orientation and sign of the pressure anisotropy in oblate plasmas. On the weak coupling side we have noted a different (but physically rather irrelevant) behavior at large quark separation: a monotonic potential in some directions and oscillatory tails reminiscent of Friedel oscillations in others.

By means of lightlike Wilson loops we have extracted anisotropic jet quenching parameters for the two holographic models and compared with perturbative leading-log calculations in hard anisotropic loops. Once again we observed qualitative changes, in particular of the ordering of $\hat q_L$ and $\hat q_\perp$ for transverse jets, when the pressure anisotropy is changed from oblate to prolate in the JW model, but not in the MT model, which is again always similar to the prolate case of the JW model, i.e.\ $\hat q_L > \hat q_\perp$. Whereas in the weak coupling calculations based on hard anisotropic loops the ordering of $\hat q_L$ and $\hat q_\perp$ depends on the sign of the pressure anisotropy, we have argued that nonperturbatively large (inhomogeneous) fields generated by plasma instabilities of a weakly coupled plasma always favor $\hat q_L > \hat q_\perp$. 

The fact that the MT model arrives at $\hat q_L > \hat q_\perp$ seems to
be related to the fact that the bulk geometry is uniformly prolate ($\mathcal H\ge1$), even when the pressure of the dual gauge theory has oblate anisotropy.
It would be very interesting to see whether dynamical holographic models of heavy-ion collisions, such as colliding shock waves in AdS \cite{Chesler:2010bi}, where both the bulk geometry and the pressure anisotropy have oblate character, would lead to different results (as the results obtained in the JW model could suggest). This seems to be a most pertinent question since heavy-ion data \cite{Putschke:2007mi} indeed point towards $\hat q_L > \hat q_\perp$, which weak coupling calculations in an anisotropic plasma appear able to reproduce \cite{Romatschke:2006bb,Baier:2008js,Dumitru:2007rp}.

\subsection*{Acknowledgments}

We are indebted to Antti Gynther, Stefan Stricker, and Aleksi Vuorinen 
for many useful discussions, and to Mike Strickland for sharing computer
code for the evaluation of heavy quark potentials in hard anisotropic loop theory.

This work was supported by the Austrian Science
Foundation FWF, project no. P22114.

\bibliographystyle{JHEP}

\bibliography{HQP}

\end{document}